# Reinforcement Learning Framework for Opportunistic Routing in WSNs

[1] G.Srinivas Rao,  [2] A.V.Ramana

[1] Department of Computer Science and Engineering, GMRIT, Rajam, AP, India

[2] Department of Computer Science and Engineering, GMRIT, Rajam, AP, India

**Abstract**

Routing packets opportunistically is an essential part of multihop ad hoc wireless sensor networks. The existing routing techniques are not adaptive opportunistic. In this paper we have proposed an adaptive opportunistic routing scheme that routes packets opportunistically in order to ensure that packet loss is avoided. Learning and routing are combined in the framework that explores the optimal routing possibilities. In this paper we implemented this Reinforced learning framework using a customer simulator. The experimental results revealed that the scheme is able to exploit the opportunistic to optimize routing of packets even though the network structure is unknown.

## 1. Introduction

Recently there has been more research into the WANETs in order to overcome the problems with traditional routing [1], [2], [3], [4]. Conventional ad hoc routing makes use of fixed path for forwarding packets [5]. However, the fixed path routing schemes suffer from taking advantage of wireless medium to have broadcasting nature of communication. The opportunistic routing in contrast to traditional routing is made in online manner. Such routing mechanisms reduce the problems with poor wireless links as they can make use of broadcast nature of wireless communications for diversity in the path selection. Markov decision is used for opportunistic routing in [1], [2], [3], [4]. This technique is used for optimal routing decision. Distance vectors can be used to know the distance and also the cost to forward details before making decisions on packet routing or forwarding. For all kinds of routing opportunistically in [4] a unifying framework is used. Other opportunistic routing techniques such as ExOR [2],Geographic Random Forwarding (GeRaF) [1] and SDF [6]. The cost measures are different in [6], [1], and [2]. A precise probabilistic model is used in all opportunistic algorithms for best routing possibilities. All such routing algorithms actually learn and maintain. To make a perfect probabilistic estimation and study an integrated approach is required. In [7] an analysis is made for opportunistic routing process [4]. However, learning in combination with opportunistic routing is not yet explored. Bhorkar et al. [8] investigated opportunistic routing of packets. Here proposed a reinforcement learning framework along with an opportunistic routing algorithm. This algorithm is expected to minimize average per-packet cost. In this paper , we achieved low-overhead, low-complexity in the implementation of the framework. However in this paper assumed no knowledge of channel statistics, use reinforcement learning framework and exploits best packet routing opportunities.

In this paper we implement the adaptive and opportunistic routing proposed by Bhorkar et al. [8]. We build a custom simulator that demonstrates the proof of concept of adaptive opportunistic routing to minimize cost of packet routing. The rest of the paper is organized into the following sections. Section 2 provides working procedure and its details. Section 2.2 shows implementation details. Section 3 provides details of experimental results while section 4 concludes the paper.

## 2. Working Procedure

This section provides the proposed scheme or transferring packets from source node to destination node. An opportunistic routing setting is assumed. We also assume a fixed transmission cost. At any given point of time only one route is capable of routing a packet. The termination condition is either the event which denotes successful packet receiving of dropping of packet. The termination time is considered as stopping time. Termination events are discriminated. The successful sending of packet to destination and the packet drop are two different things. However, both are considered to be termination events. The nodes are expected to participate the packet routing genuinely as they are given average per-packet reward. Such reward is computed as follow.

$$J_N = E\left[\frac{1}{M_N}\sum_{m=1}^{M_N}\{r_m - \sum_{n=r_s^m}^{t_T^m-1} c_{i_{n,m}}\}\right] \quad (1)$$

here, $i_{n,m}$ indicates index of node at time n when sends packet m and $c_{i_{n,m}}$ indicates cost measure. It will become zero when no packets are transmitted at time n. As per above equation, the routing scheme of nodes can be seen for relaying packets m. In the above equation $M_N$ indicates number of packets terminated up to time n. In this model we are choosing relay of nodes $i_{n,m}$ in the absence of knowledge about network structure in a way that $J_N$ is to be maximum when N is increasing. to solve this issue we proposed distributed algorithm and explained in section 2.1.

2.1 Distributed Algorithm

The Reinforced learning framework for opportunistic routing scheme uses a distributed algorithm. The algorithm uses various notations. The notations and the meanings of the notations are presented in table 1.

**Table 1**: Notations used in algorithm

| Symbol | Definition |
|---|---|
| $S_n^i$ | Set of nodes which receives transmissions at time n from node i |
| $a_n^i$ | Node *i* taken decision at time n |
| A(S) | Set of available actions when nodes in S receive a packet |
| N(i) | Neighbors actions when nodes in S receive a packet. |
| g(S,a) | R obtained by taking decision a when set S of nodes receive a packet. |
| $v_n$(i,S,a) | upto time n, number of times the nodes in S received a packet from node i and decision *a* is taken |
| $N_n$(i,S) | Number of times upto time n, nodes S have received a packet from node i |
| $^\wedge_n$(i,S,a) | Number of times upto time n, nodes S have received a packet from node *i* and *a* is taken |
| $\Lambda_{max}^i$ | Estimated best score for node i |

The algorithm flow is as shown in Fig. 1. The algorithm has four stages namely transmission, acknowledgement, relay and update. They are visualized in the following figure

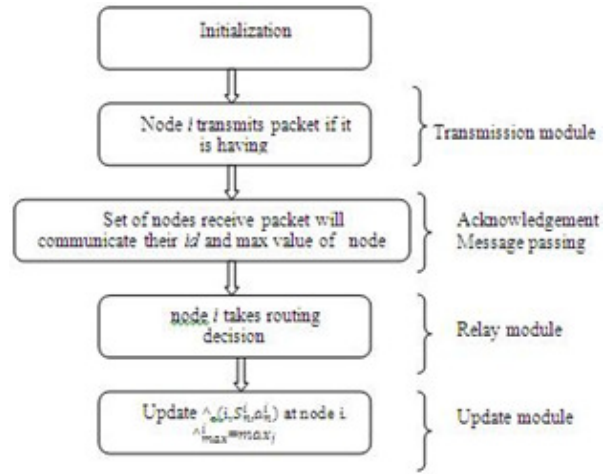

Fig.1 – Phases of distributed algorithm

As can be seen in fig. 1, the initialization phase initializes required parameters like $^\wedge_0(i,S,a)=0$, $v_{-1}(i,S,a)=0$, $\Lambda_{max}^i=0$; the transmission phase is used to transmit packets when a node is having packet; the acknowledgement phase is meant for acknowledge the receiving of packet. Here, set of nodes which receives packet will communicate their id and max score ; relay phase is used to make routing decision; and update phase updates the parameters, here updates $^\wedge_n(i,S_n^i,a_n^i)$ at node i. More details can be found in [8]. routing decision at any given time is made based on the reception outcome and involves retransmission, choosing the next relay, or termination. Our proposed scheme makes such decisions in a distributed manner via the following handshaking mode between node *i* and neighbor nodes.

1) At particular time *n*, node *i* transmits a packet.
2) Set of nodes S who successfully received the packet from node *i*, transmit acknowledgment (ACK) packets to node *i*. Apart from node's identity, the acknowledgment packet of node includes a control *message* known as *estimated best score* (EBS).
3) Node *i* announces j which is belongs to S as the next transmitter or announces the termination decision T in a forwarding (FO) packet.

Node *i* takes routing decision based on the scare vector at particular time n .by using estimated best score, the score vector of nodes is updated.

2.2 Implementation of Custom Simulator

The customer simulator is implemented using Java programming language. It is based on the adaptive opportunistic routing scheme proposed by Bhorkar et al.

[8]. The environment used is a PC with 4GB RAM and Core 2 Duo processor. Net Beans is used for rapid application development. SWING API in Java is used to build nodes in the network. The application is tested with plethora of nodes. Each node interface appears as follows. As can be seen in fig. 2 the user interface of a node has two tabs. The first tab provides node related details while the second node is related to the paths established as per the proposed adaptive opportunistic routing algorithm. As can be seen in fig. 3 the user interface of a node has two tabs. The first tab provides node related details while the second node is related to the paths established as per the proposed adaptive opportunistic routing algorithm. The experimental results through the nodes indicated that the simulator is capable of demonstrating him proof of concept. The following section provides experimental results.

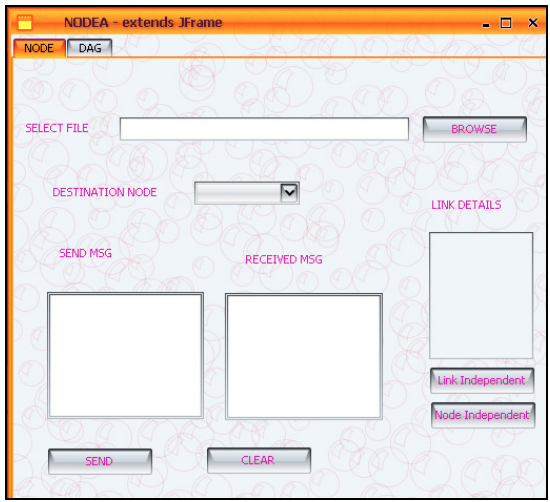

Fig. 2 Illustrates a node in the network (Node A)

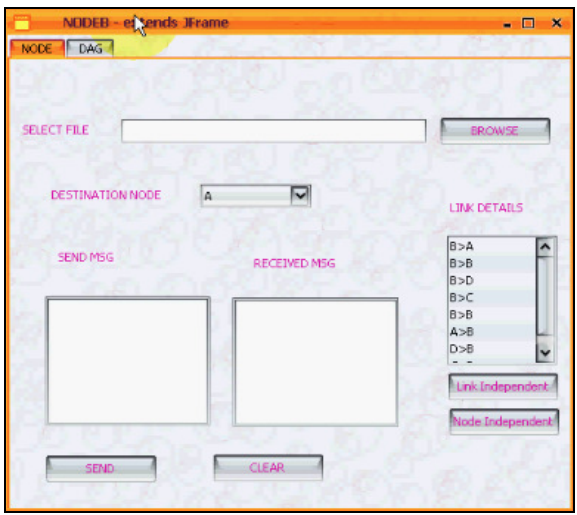

Fig. 3 Illustrates a node in the network (Node B)

## 3. Experimental Results

The experimental results that have been obtained the simulations done through custom simulator are presented in this section. When a node is ready to send a packet, it will send through best path by calculations best scores and keeping record of vector score for that node. in future transactions, it will use the score for best path finding.

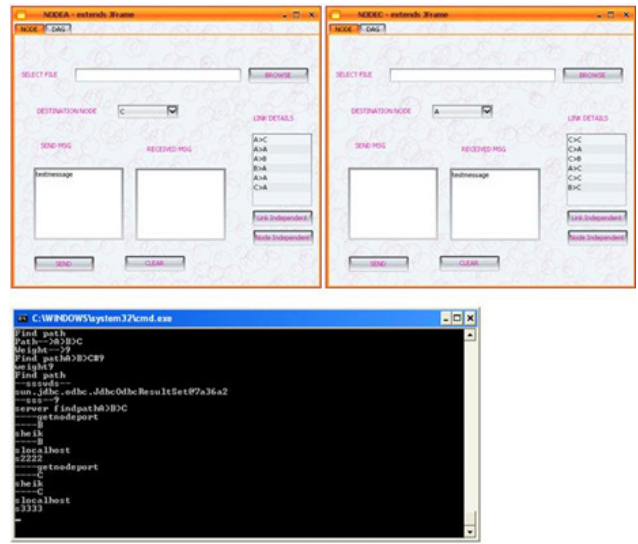

Fig. 4-Routing of packets in adhoc networks when there is no traffic.

When a particular node is not Wireless network, the system will route the packets by using another paths and results pertaining to the kind of behavior is shown below.

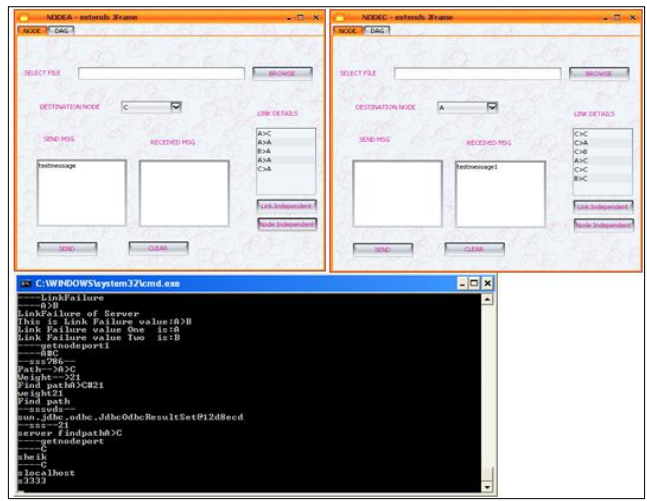

Fig.5-Scenario of routing of packets when a node disconnected in network.

We can show the transmissions per packet vs time as Graph as mentioned below.

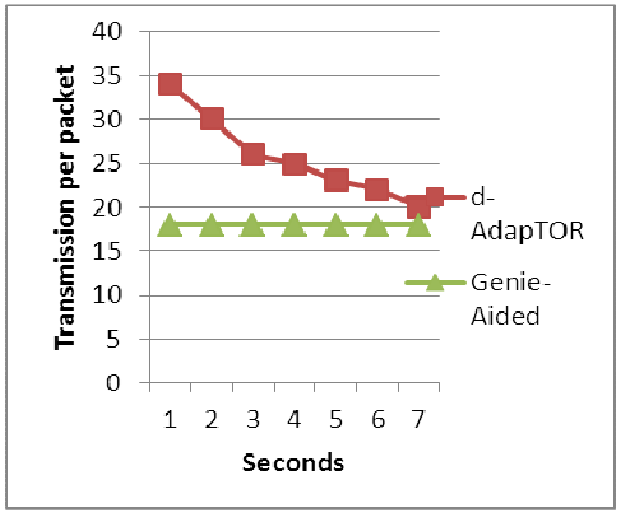

Fig. 6 Transmissions per packet vs. seconds.

As can be seen in fig. 6, horizontal axis represents time in seconds, while the vertical axis represents transmissions per packet. From the graph it can be understood that as the time goes the transmissions per packet is lost. We are comparing the performance of d-AdaptOR against the performance of genie-aided policy that relies on full network topology information when selecting routes.

As can be seen in fig. 7, horizontal axis represents R value which represents delivery reward, while the vertical axis represents expected delivery ratio. From the graph it can be understood that as the R value increases, the expected delivery ratio increases. A reasonable choice of R is any value larger than the worst-case expected transmission cost. Increasing beyond such a value does not affect the asymptotic optimality of the algorithm.

Fig. 7 plots the delivery ratio as R is varied. Fig. 7 shows that as increases beyond a particular level , the delivery ratio remains fixed. However, for sufficiently small , nearly all the packets are dropped as the cost of transmission of the packet as well as relaying is not worth the obtained delivery reward.

## 4. Conclusion

In this paper we implemented Adaptive Opportunistic routing proposed by Bhorkar et al. [8] using a Java custom simulator. Though zero knowledge is assumed about the channel statistics and topology, the scheme is capable of exploring and exploiting adaptive opportunistic routing possibilities in order to reduce the average per-packet cost. This is achieved by providing rewards to the nodes that perform well in packet routing. The proposed simulator is capable of demonstrating the proof of concept with respect to opportunistic routing. The empirical results revealed that the distributed nature of algorithm along with reward system has produced best results in routing of packets.

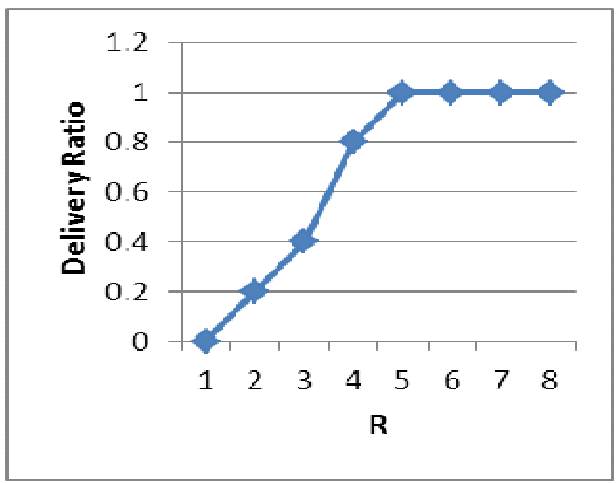

Fig. 7 Delivery ratio vs. R

**G.Srinivas Rao** received his B Tech in Computer Science and Engineering from Jawaharlal Nehru Technological University,

Hyderabad, A.P, INDIA. in 2010 and pursing M.Tech in GMRIT, Rajam, A.P, India. His area of interest is Wireless Networks.

**A.V. Ramana** obtained the MCA Degree from Andhra University ,Visakhapatnam ,A.P ,India. Obtained M.E  degree from Anna Univeristy . He is presently working as  Assistant Professor in GMR Institute of Technology, Rajam, AP, India. His area of interest is Mobile Networks and wireless Networks.